\documentclass{elsart}
\usepackage{cases}
\usepackage{epsfig}
\newlength{\wordlength}

\newlength{\onewordlength}

    \newcommand{\ba}{\begin{eqnarray}}
    \newcommand{\ea}{\end{eqnarray}}
    \newcommand{\be}{\begin{equation}}
    \newcommand{\ee}{\end{equation}}

    \newcommand{\AmS}{{\protect\the\textfont2%
  A\kern-.1667em\lower.5ex\hbox{M}\kern-.125emS}}

\newcommand {\bk} {{\mathbf k}}
\newcommand {\bq} {{\mathbf q}}

\newcommand {\bp} {{\mathbf p}}

\newcommand{\bx}{{\bf x}}

\newcommand{\calF}{{\mathcal F}}

\newcommand{\calP}{{\mathcal P}}

\begin{document}
\runauthor{PKU}
\begin{frontmatter}

\title{Volume Dependence of Spectral Weights for Unstable Particles in a Solvable Model}
 \author[PKU]{Guozhan Meng}
\author[PKU]{Chuan Liu}

\address[PKU]{School of Physics and Center for High Energy Physics\\
          Peking University\\
                  Beijing, 100871, P.~R.~China}
                \thanks{This work is supported by the National Natural
 Science Foundation (NFS) of China under grant
 No. 10721063, No. 10675005 and No. 10835002.}

 \begin{abstract}
 Volume dependence of the spectral weight is usually
 used as a simple criteria to distinguish single-particle states from multi-particle
 states in lattice QCD calculations. Within a solvable model, the Lee model,
 we show that this criteria is in principle only valid
 for a stable particle or a narrow resonance. If the resonance being
 studied is broad, then the volume dependence of the
 corresponding spectral weight resembles that of a multi-particle
 state instead of a single-particle one.
 For an unstable $V$-particle in the Lee model,
 the transition from single-particle to multi-particle volume
 dependence is governed by the ratio of its physical width to
 the typical level spacing in the finite volume.
 We estimate this ratio for practical lattice QCD simulations
 and find that, for most cases, the resonance studied in lattice
 QCD simulations still resembles the single particle behavior.
 \end{abstract}
 \end{frontmatter}


 \section{Introduction}

 Quantum Chromodynamics (QCD) is believed to be the underlying theory of strong interactions.
 Due to its non-perturbative nature, low-energy properties of strong interaction
 should be studied with a non-perturbative method. Typical problems
 include light hadron spectrum and low-energy hadron-hadron scattering.
 Lattice QCD provides a genuine non-perturbative framework in which
 non-perturbative problems can be tackled using numerical simulations.
 In a typical lattice calculation for hadron spectrum,
 energy eigenvalues of the QCD Hamiltonian is measured numerically, with different
 quantum numbers that are conserved by the strong interaction.
 \footnote{Strictly speaking, only the eigenvalues at finite
 lattice spacing are measured. To obtain the continuum eigenvalues,
 one has to perform the continuum extrapolation of lattice results.}
 People tend to interpret these energy eigenvalues
 as mass values of corresponding particles. This seems to provide
 a non-perturbative definition for the mass value of a hadron.
 The width of a hadronic resonance is a more complex issue.
 Using L\"uscher's formula, scattering phase shifts
 can also be calculated from the two-particle energy
 eigenvalues~\cite{luscher86:finiteb,luscher90:finite,luscher91:finitea,luscher91:finiteb,%
chuan04:asymmetric,chuan04:asymmetric_long,chuan05:2channel,%
CPPACS03:pipi_phase,CPPACS04:pipi_phase_unquench,savage06:pipi,%
savage06:NN,savage06:piK}.
 However, a direct, model-independent and non-perturbative calculation of the width
 parameter remains a difficult
 task~\cite{michael05:decay,michael06:hybrid_decay,fiebig06:hybrid_decay,CPPACS07:rho_decay}.

 Phenomenologically, a resonance is characterized by
 its mass parameter $M$ and the width parameter $\Gamma$.
 A common theoretical definition for these two physical parameters refers
 to the pole of the $S$-matrix on the second-sheet of the complex energy plane:
 \be
 \label{eq:def_pole}
 z=M^{(p)}-i\Gamma^{(p)}/2\;,
 \;\;  S(z)\rightarrow \infty\;.
 \ee
 However, experimentalists prefer more tractable definitions
 such as the scattering phase shift, or the total cross-sections which are in principle measurable
 physical quantities in the scattering experiment.
 For example, the position for the mass of a resonance
 can be defined to be the position
 where total cross section reaches its maximum or the
 corresponding phase shift passing $\pi/2$.
 \footnote{In this discussion, we have neglected all
 other background contributions. We also assume that
 in the energy range we are investigating, there is only one resonance.}
 The definition of the width in this case is somewhat ambiguous
 except for narrow resonances. In the case of
 infinitely narrow resonance, when
 $E_{c.m.}=M^{(\delta)}\pm\Gamma^{(\delta)}/2$, the phase shift exactly
 passes through $\pi/4$ and $3\pi/4$, respectively.
 In terms of total cross-section, this also corresponds to
 the position where cross-section has dropped to half
 its peak value. For wide resonances, the peak is usually not
 symmetric with respect to the peak position and we
 may choose to define the width by demanding the
 phase shift to be exactly $\pi/4$ at
 $M^{(\delta)}-\Gamma^{(\delta)}/2$. So, using the phase shift,
 one possible definition goes:
 \be
 \label{eq:def_delta}
 \delta(M^{(\delta)})=\pi/2\;,
 \delta(M^{(\delta)}-\Gamma^{(\delta)}/2)=\pi/4\;.
 \ee
 It is a well-known fact that, the above mentioned definitions
 for the spectral parameters of a resonance, namely the energy eigenvalues
 measured in lattice calculations, the $S$-matrix pole definition and
 the phase-shift definition,  do not coincide with
 one another in general. One expects that they only agree  when the resonance
 becomes infinitely narrow. Although it is difficult to show this
 for a general theory non-perturbatively, we will show this explicitly in
 a totally solvable field theoretical model, the Lee model.

 Numerical simulations in lattice QCD are performed within a finite
 volume. All energy levels in this finite box are discrete.
 Therefore, it is an important and  non-trivial question for the lattice
 calculations to properly identify single-particle states and
 multi-particle states which might mix within a particular symmetry
 channel. A typical example is the $\rho$ meson, which is a single resonance that mixes
 with two pion scattering states.
 To distinguish the single-particle states from multi-particle states, it is suggested that
 spectral weights of the eigenstates are to be measured. One expects
 that spectral weights of single and multi-particle eigenstates
 show different volume dependence and thus can be utilized to
 disentangle the single-particle states from the multi-particle
 states. Typically, the spectral weight of a single-particle state has little
 volume dependence if the volume is not too small while the spectral weight
 of two-particle states will exhibit a typical $1/\Omega$ dependence
 which can be captured by performing the simulation on two different
 $\Omega$ which is the three-volume of the lattice.
  This strategy has been used in Ref.~\cite{KFLiu04:pentaquark} where the authors show that
  the so-called penta-quark states measured in their lattice calculations are
  in fact kaon-nucleon two-particle scattering states.  However, this conclusion
  is not so settled even in the first-principle lattice QCD
  calculations~\cite{Takahashi:2005uk,Csikor:2005xb,Ishii:2004qe}.

 Although well-suited for a stable particle or a narrow resonance, one
 also expects the above mentioned criteria to be modified when the resonance
 becomes broad. The reason for this is quite clear.  For a broad
 resonance, the scattering states themselves form a complete set
 in the corresponding Hilbert space and
 one thus expects the spectral weight of a broad resonance to
 behave more like that of multi-particle scattering states.
 Although this sounds reasonable, the transition of the spectral weight from
 the single-particle behavior to
 the multi-particle behavior has never been shown explicitly
 in the literature. In this paper, we demonstrate this
 scenario within the Lee model where everything can be computed explicitly.
 This result suggests that, using the volume dependence of spectral weight
 as {\em the criteria} to distinguish single-particle from the
 multi-particle states is only valid when the
 single particle is either stable or unstable but narrow.
 In other words, it cannot be applied to broad
 resonances without further careful considerations.
 Moreover, within the Lee model we can also verify that,
 the transition from single-particle to multi-particle
 behavior is governed by the ratio of the physical width
 to the typical level spacing in the
 finite box. Assuming this ratio is also the relevant quantity
 in lattice QCD, we estimate this ratio for some typical lattice
 calculations and find that, in most cases, the resonance studied
 still resembles the single particle behavior for recent lattice
 simulations and conclude.

 This paper is organized as follows. In section 2, we introduce
 the Lee model and summarize its main results.
 The scattering phase shifts and the $S$-matrix element are
 also calculated and various definitions for the spectral parameters
 of a resonance is compared directly. In section 3, we focus
 on the Euclidean correlation functions that are measured in
 lattice simulations. Spectral weights for the eigenstates are
 computed and the volume dependence of the spectral weights are
 analyzed. As the resonance becomes broader, the transition from
 the single-particle behavior to the multi-particle behavior is
 explicitly shown. In section 4, we will discuss possible impact of our results
 by estimating the ratio in recent lattice simulations on
 pion-pion scattering.

 \section{The energy eigenstates and the phase shift in the Lee model}
 \label{sec:model}

 The Lee model~\cite{Lee54:model} is a
 completely solvable field theoretical
 model proposed by Lee long time ago. The model involves
 three types of ``particles": the so-called $V$-particle,
 the $N$-particle and the $\theta$-particle.
 The Hamiltonian of the model is given by:
 \ba
 \label{eq:lee_model_H}
 H&=&H_0+H_1\;, \nonumber \\
 H_0 &=&m_V\sum_{\bp}V^\dagger_\bp V_\bp
    +m_N\sum_{\bp}N^\dagger_\bp N_\bp
    +\sum_{\bk}\omega_\bk a^\dagger_\bk a_\bk\;,
    \nonumber \\
 H_1 &=&-{g_0\over\sqrt{\Omega}}\sum_{\bp,\bk}
 {f(\omega_\bk)\over\sqrt{2\omega_\bk}}
 \left[V^\dagger_{\bp+\bk}N_\bp a_\bk+
 V_{\bp+\bk}N^\dagger_\bp a^\dagger_\bk\right]
 \;.
 \ea
 Here $\Omega$ is a large but finite volume of the system;
 $g_0$ is the bare coupling constant.
 $V^\dagger_\bp$ ($V_\bp$), $N^\dagger_\bp$ ($N_\bp$) and
 $a^\dagger_\bp$ ($a_\bp$) corresponds to the creation
 (annihilation) operators of the $V$, $N$ and $\theta$-particles,
 respectively. They satisfy the usual commutation relations:
 \footnote{In the original Lee model, the $V$-particles and
 $N$-particles are fermions while $\theta$ particles are bosons.
 In this letter, we assume they are all bosons. This does not
 change the results.}
 \be
 [V_\bp, V^\dagger_\bk]=
 [N_\bp, N^\dagger_\bk]=
 [a_\bp, a^\dagger_\bk]=\delta_{\bp\bk}.
 \ee
 We will call $m_V$, $m_N$ in~(\ref{eq:lee_model_H}) bare
 mass of the $V$ and $N$ particle. The energy of the
 $\theta$ particle is given by:
 $\omega_\bk=\sqrt{\mu^2+\bk^2}$
 where $\mu$ is the mass of the $\theta$-particle.
 To make the whole system well-defined, we have enclosed
 the system in a three-volume of size $\Omega$ and introduced
 a form factor $f(\omega)$ to regularize the possible ultra-violet
 divergences. It is easy to see that the free one $N$-particle and one
 $\theta$-particle states remain eigenstates of the full
 Hamiltonian. However, the free one $V$-particle states
 are not since it is coupled to the $N\theta$ pair states.
 We will restrict our discussion in the sector
 of one $V$-particle and $N\theta$ pair states.

 We are concerned with the following properties
 of the Lee model: the exact eigenstates and eigenvalues
 of the Hamiltonian, which is what is measured in lattice
 simulations; the exact $S$-matrix element and scattering phase
 shift, which will be utilized to locate possible resonances in
 the model, and the spectral weight from the Euclidean correlation
 function whose volume dependence is our major concern in this paper.
 In this section, we will first summarize the results for the
 eigenstates, eigenvalues and $S$-matrix elements. Spectral weights
 will be dealt with in the next section.

 In the $V$-$N\theta$ sector, the exact eigenstates $|n\rangle_\bp$
 of the full Hamiltonian (with eigenvalue $E_n$) can be obtained
 as~\cite{Glaser55:unstable}:
 \ba
 \label{eq:eigenstates}
 |n\rangle_\bp &=& Z^{-1/2}_n\left[
 |V_\bp\rangle+{g_0\over\sqrt{\Omega}}\sum_\bk
 {f(\omega_\bk)\over\sqrt{2\omega_\bk}}
 {1\over m_N+\omega_\bk-E_n}
 |N_{\bp-\bk}\theta_\bk\rangle
 \right]\;, \\
 \label{eq:eigenvalues}
 m_V-E_n &=& F(E_n-m_N)\;, \;\;
 F(x) \equiv {g^2_0\over\Omega}\sum_\bk
 {f^2(\omega_\bk)\over 2\omega_\bk}
 \left({1\over \omega_\bk-x}\right)\;,
 \ea
 and the normalization factor $Z_n$ is also found to be:
 \be
 \label{eq:Z}
 Z_n(E_n)=1+F'(E_n-m_N)=1+ {g^2_0\over\Omega}\sum_\bk
 {f^2(\omega_\bk)\over 2\omega_\bk}
 \left({1\over \omega_\bk+m_N-E_n}\right)^2\;.
 \ee
 The function $F(x-m_N)$ has
 simple poles at each $x=m_N+\omega_\bk$ and
 one obtains a series of eigenvalues $E_n$ from
 Eq.~(\ref{eq:eigenvalues}).
 It is also seen that there is always a root
 $E_n$ satisfying $E_n<m_N+\mu$. However, one cannot
 draw the conclusion that the $V$-particle is
 always stable. The fact is that, if $m_V$ is small
 enough, then a stable $V$ particle exists. If
 $m_V$ is too large, then no stable $V$-particle
 exists. The precise condition in the infinite
 volume limit is~\cite{Glaser55:unstable}:
 \be
 \label{eq:unstable}
 m_V-m_N >\mu +\phi(\mu)\;,
 \ee
 where $\phi(x)$ is the principle-valued
 integral counterpart of the function $F$
 defined in Eq.~(\ref{eq:eigenvalues}):
 \be
 \label{eq:principle_value}
 \phi(x)=g^2_0\int {d^3k\over (2\pi)^3}
 {f^2(\omega_\bk)\over 2\omega_\bk}
 \calP\left({1\over \omega_\bk-x}\right)\;,
 \ee
 Note that the function $\phi(x)$ is a monotonically increasing function of $x$.
 Therefore, if condition~(\ref{eq:unstable}) is
 satisfied, $V$-particles are unstable and they
 decay into $N\theta$ particle pairs.

 At this stage, it is useful to point out the following fact.
 If we were to replace the eigenvalues $E_n$ in Eq.~(\ref{eq:eigenstates})
 by $E+i\lambda$ with both $E$ and $\lambda$ being real but
 $\lambda$ is small, we can construct an
 {\em approximate} eigenstate
 of the Hamiltonian. We find that the value of $E$ has to be one of those
 $E_n$ values. However, it is easy to see that the equation for
 the imaginary part $\lambda$ can never be satisfied exactly
 for non-vanishing $\lambda$.
 This is due to the fact that the Hamiltonian is Hermitian
 and the eigenvalues thus have to be real. However, if we take
 $E$ to be one of those $E_n$ values but $\lambda$ being small enough,
 we indeed obtain an approximate eigenstate of the Hamiltonian.
 The approximation becomes better and better as
 $\lambda\rightarrow 0$. It is well-known that a narrow resonance
 in scattering theory in fact corresponds to such a scenario.
 Nevertheless, one should keep in mind that such a description
 is in fact only meaningful when the resonance is narrow.

 Not only the exact energy eigenstates can be obtained,
 the scattering phase shifts can also be calculated within
 this sector of the Lee model. It can be shown that,
 when the $V$ particle is unstable, the $N\theta$ scattering
 states form a complete set in the Hilbert space. These scattering
 states are the solution of the corresponding Lippmann-Schwinger equation:
 \be
 \label{eq:LS}
 |N_\bq\theta_\bk\rangle_\pm
 =|N_\bq\theta_\bk\rangle + {1\over m_N+\omega_\bk-H\pm
 i\epsilon}H_1|N_\bq\theta_\bk\rangle\;.
 \ee
 The states $|N_\bq\theta_\bk\rangle_{+/-}$ corresponds to
 well-prescribed incoming/outgoing waves in the
 infinite past/future, respectively.
 These states are also eigenstates of the full Hamiltonian
 with eigenvalues $m_N+\omega_\bk$.
 It can be shown explicitly that both $|N_\bq\theta_\bk\rangle_+$
 and $|N_\bq\theta_\bk\rangle_-$ form a complete set in this
 particular Hilbert subspace. They also form an ortho-normal basis:
 \be
 {}_{\pm}\langle N_{\bq'}\theta_{\bk'}|N_\bq\theta_\bk\rangle_\pm
 =\delta_{\bq\bq'}\delta_{\bk\bk'}\;.
 \ee
 The unitary matrix which relates these two sets of
 ortho-normal states is nothing but the $S$-matrix whose
 matrix elements is defined via:
 \be
 S_{\bq'\bk';\bq\bk} \equiv
 {}_{-}\langle N_{\bq'}\theta_{\bk'}|N_\bq\theta_\bk\rangle_+
 \;.
 \ee
 For the Lee model, the Lippmann-Schwinger states defined
 in Eq.~(\ref{eq:LS}) can be computed exactly with
 the result~\cite{Glaser55:unstable,schweber:qft}:
 \ba
 \label{eq:solution}
 |N_\bq\theta_\bk\rangle_\pm &=& |N_\bq\theta_\bk\rangle
 -{g_0f(\omega_\bk)\over\sqrt{2\Omega\omega_\bk}}
 {1\over m_N+\omega_\bk -m_V \pm i\epsilon +F(\omega_\bk\pm i\epsilon)}
 \nonumber \\
 &\times&
 \left[ |V_{\bk+\bq}\rangle -{g_0\over\sqrt{\Omega}}\sum_\bp
 {f(\omega_\bp)\over\sqrt{2\omega_\bp}}
 {1 \over\omega_\bk-\omega_\bp\pm i\epsilon}
 |N_{\bq+\bk-\bp}\theta_\bp\rangle\right]\;.
 \ea
 From this result, one gets the scattering phase shift
 for the $N\theta$ scattering. It turns out that there is
 only $s$-wave scattering in this sector of the Lee model
 and the corresponding phase shift satisfies the following equation:
 \ba
 \label{eq:phase_shift}
 e^{2i\delta(k)} &=& {m_N+\omega-m_V+\phi(\omega)-i\Gamma(\omega)/2
 \over m_N+\omega-m_V+\phi(\omega)+i\Gamma(\omega)/2}\;,
 \nonumber \\
 \tan\delta(k) &=& -{\Gamma(\omega)/2\over
 m_N+\omega-m_V+\phi(\omega)}\;,
 \ea
 where $\omega=\sqrt{k^2+\mu^2}$ and the function
 $\phi(\omega)$ is given by Eq.~(\ref{eq:principle_value}).
 If the parameters of the theory is chosen such that
 $m_N+\mu-m_V+\phi(\mu)<0$ at the threshold,
 which is the same condition for the $V$-particle becoming unstable,
 it is seen that the real part of the denominator
 in the $S$-matrix will vanish at
 some energy $M^{(\delta)}$ above the threshold.
 At this particular energy,
 the model exhibits a typical resonance. However, the
 (bare) width of the resonance is in general not a constant
 but energy dependent:
 \be
 \label{eq:width}
 \Gamma(\omega)={g^2_0\over 2\pi}f^2(\omega)\sqrt{\omega^2-\mu^2}
 \theta(\omega-\mu) \;.
 \ee
 The total energy at which the phase shift passes through $\pi/2$
 is given by:
 \be
 M^{(\delta)}-m_V+\phi(M^{(\delta)}-m_N)=0\;,
 \ee
 while the definition for the width gives:
 \ba
 M^{(\delta)}-{\Gamma^{(\delta)}\over 2}-m_V
 &+&\phi\left(M^{(\delta)}-{\Gamma^{(\delta)}\over 2}-m_N\right)
 \nonumber \\
 &+&{1\over 2}\Gamma\left(M^{(\delta)}-{\Gamma^{(\delta)}\over 2}-m_N\right)=0
 \;.
 \ea
 The pole mass and the corresponding width are given by:
 \be
 z^{(p)}=M^{(p)}-i{\Gamma^{(p)}\over 2}\;,
 \;\;
 z^{(p)}-m_V+\calF(z^{(p)}-m_N)=0
 \;,
 \ee
 where the function $\calF(z)$ is given by:
 \be
 \calF(z)=\int^\infty_\mu{\Gamma(\omega)\over 2\pi}
 {d\omega \over \omega-z}\;,
 \ee
 with the understanding that this pole position
 should be solved on the second sheet.
 \footnote{It can be shown that the above equation
 can never be satisfied on the first sheet where
 the solutions correspond to stable bound states on
 the real axis.}
 Comparison of the above explicit formulae shows that they are
 generally different if the width of the resonance is not narrow.
 It is also evident from the above formulae that
 when the width is becoming infinitely narrow,
 $M^{(\delta)}$ coincides with $M^{(p)}$ while
 $\Gamma^{(\delta)}$ coincides with $\Gamma^{(p)}$.

 \section{The Euclidean correlation functions and the spectral weights}
 \label{sec:euclidean}

 In this section, we  discuss the mass values and the corresponding
 spectral weights measured in a lattice Monte Carlo simulation.
 In such a calculation, by measuring appropriate Euclidean correlation
 functions, the eigenvalues (typically a few lowest)
 of the Hamiltonian is obtained.
 \footnote{In reality, lattice data still contain lattice
 artifacts caused by the finite lattice spacing. In this paper,
 we assume that these finite lattice spacing errors have already
 been subtracted, namely the continuum limit is already taken.}
 In the Lee model, these
 eigenvalues are precisely those $E_n$ values given by Eq.~(\ref{eq:eigenvalues}).
 It is then clear that $E_n$ in principle is different
 from any of $M^{(p)}$ or $M^{(\delta)}$ defined by the
 $S$-matrix pole or the phase shift. But if the resonance
 is narrow enough, $E_n$ coincides with $M^{(p)}$ or $M^{(\delta)}$.
 As pointed out in the introduction, it is suggested
 that one can distinguish the single-particle,
 two-particle and multi-particle states
 by inspecting the volume dependence of the so-called
 spectral weights for the states~\cite{KFLiu04:pentaquark}.
 Here we would like to investigate this possibility within the Lee model where the eigenstates
 and the corresponding eigenvalues are explicitly known.

 We will first look at an interpolating field $V(\bx)$.
 The correlation function that we are interested in is:
 \be
 \sum_\bx \langle 0|V(\bx,t)V^\dagger(0)|0\rangle\;,
 \ee
 where we have assumed that the fields are now defined
 on a lattice with the lattice spacing being set to unity.
 Inserting the complete set of states we have:
 \be
 \label{eq:spectral_weight}
 \sum_\bx \langle 0|V(\bx,t)V^\dagger(0)|0\rangle
 \propto \sum_n Z^{-1}_n(E_n) e^{-E_nt}\;.
 \ee
 where $E_n$ and $Z_n(E_n)$ are given by
 Eq.~(\ref{eq:eigenvalues}) and Eq.~(\ref{eq:Z}) respectively.
 Therefore, the spectral weight function $W_n$ for each
 eigenstate $|n\rangle_\bk$ is simply:
 \be
  \label{eq:spectral_weight1}
 W_n=Z^{-1}_n(E_n)=\left(1+{g^2_0\over\Omega}\sum_\bk
 {f^2(\omega_\bk)\over 2\omega_\bk}
 \left({1\over \omega_\bk+m_N-E_n}\right)^2
 \right)^{-1}\;.
 \ee
 At first sight, the spectral weights in
 Eq~(\ref{eq:spectral_weight}) and Eq~(\ref{eq:spectral_weight1})
 do not seem to show the expected volume dependence at all.
 However, we will show below that if the $V$-particle is stable or
 if the width of the unstable $V$-particle is small,
 Eq.~(\ref{eq:spectral_weight1}) does provide the expected volume
 dependence for single and two-particle states respectively.

 In general, the volume dependence of $W_{n}$
 is quite complicated for a finite (not necessarily large)
 volume $\Omega$. However, if the volume $\Omega$ is sufficiently
 large, the volume dependence can be estimated.
 As Eq.~(\ref{eq:spectral_weight1}) shows,
 one is led to consider the function $F(x)$
 defined in Eq.~(\ref{eq:eigenvalues}).
 The spectral weight of a particular energy eigenvalue
 is simply related to the derivative of this function
 evaluated at the exact energy eigenvalue:
 \be
 W_n=1/(1+F'(E_n-m_N))\;.
 \ee

 The behavior of the function $F(x)$ is drastically
 different for values of $x$ below the
 threshold and above the threshold.
 If $x$ is below the threshold, i.e. $x<\mu$,
 the contribution to be summed is bounded
 in the large volume limit and the function
 $F(x)$ goes over to its integration
 counterpart $\phi(x)$ smoothly as the
 volume goes to infinity.
 However, if $x$ is
 above the threshold ($x>\mu$), there exist
 values of $\omega_\bk$ which are sufficiently close
 to $x$ in the large volume limit and therefore
 some contributions are unbounded. We will discuss
 this situation in the following.

 For large enough three volume $\Omega$, a typical
 spacing between adjacent energy levels, which we
 denote as $\Delta\omega$, can be estimated as follows:
 \be
 \label{eq:level-spacing}
 {\Omega\over (2\pi)^3}g(\omega)\Delta\omega=1\;,
 \mapsto
 \Delta\omega ={(2\pi)^3\over\Omega}{1\over g(\omega)}\;,
 \ee
 where $g(\omega)=4\pi\sqrt{\omega^2-\mu^2}\omega$ is the
 density of states for the $N\theta$ pairs. Therefore,
 in the infinite volume limit, the level spacing is
 proportional to $1/\Omega$.

 In the definition of $F(x)$, the function to be summed over
 factorizes into two parts: the fast-changing part $1/(\omega-x)$
 and the slow-changing part $f^2(\omega)/(2\omega)$.
 Here the term slow-changing refers to the fact
 that when $x$ changes an amount of the order of $\Delta\omega$,
 the function changes little (and likewise for the
 definition of fast-changing).
 Note that this factorization is meaningful
 only when the volume is large and hence $\Delta\omega$ is
 small. Assuming that we are in such a situation, then
 the summation for the function $F(x)$ may be separated
 into two parts:
 \be
 F(x) = {g^2_0\over\Omega}
 \left(\sum_{\bk,|\omega_\bk-x|\ge\epsilon}
 +\sum_{\bk,|\omega_\bk-x|<\epsilon}\right)
 {f^2(\omega_\bk)\over 2\omega_\bk}
 \left({1\over \omega_\bk-x}\right)\;,
 \ee
 where $\epsilon$ is a small positive number within which
 the function $f^2(\omega)/(2\omega)$ is almost a
 constant, but $\epsilon \gg \Delta\omega$.
 The first summation in the above expression is
 nothing but the principle-valued integral
 $\phi(x)$ once the volume is going to infinity
 and the parameter $\epsilon$ is going to zero.
 We will denote it as: $\phi_\epsilon(x)$. In the
 second summation, since the function
 $f^2(\omega)/(2\omega)$ can be viewed as a
 constant, we have:
 \be
 F(x)=\phi_\epsilon(x)
 +{g^2_0\over\Omega}
 {f^2(x)\over 2x}
 \sum_{\bk,|\omega_\bk-x|<\epsilon}
 {1\over \omega_\bk-x}\;.
 \ee
 Now that the density of state function $g(\omega)$ is
 also a slow-changing function of the energy, therefore,
 within the interval $|\omega_\bk-x|<\epsilon$, the
 energy levels can be viewed as almost equally-spaced
 with the level spacing given by Eq.~(\ref{eq:level-spacing}).
 Denoting the level closest to $x$ by $\omega^*$, we have:
 \be
 \sum_{\bk,|\omega_\bk-x|<\epsilon}
 {1\over \omega_\bk-x}
 \simeq \sum^\infty_{n=-\infty}{1\over \omega^*+n\Delta\omega-x}
 \;,
 \ee
 where we have extended the summation to infinity.
 Now the summation can be computed exactly and using
 the relation in Eq.~(\ref{eq:level-spacing})
 and the definition~(\ref{eq:width}) we
 finally have:
 \makeatletter
 \let\@@@alph\@alph
 \def\@alph#1{\ifcase#1\or \or $'$\or $''$\fi}\makeatother
 \begin{subnumcases}
 {F(x)=}
 \phi(x), &$x< \mu$, \label{eq:Fbelow}\\
 \phi(x)-{\Gamma(x)\over 2}
 \cot\left[\pi\left({x-\omega^*\over\Delta\omega}\right)\right], &$x\ge \mu$.
 \label{eq:Fabove}
 \end{subnumcases}
 \makeatletter\let\@alph\@@@alph\makeatother
 This expression is a good estimate for the function $F(x)$
 in the large volume limit for $x>\mu$.
 Note that if we set $x-(m_V-m_N)+F(x)=0$, which is nothing
 but the eigenvalue equation~(\ref{eq:eigenvalues}), we
 would obtain all the energy eigenvalues: $x=E_n-m_N$.
 Using the estimate~(\ref{eq:Fabove}) and the
 result for the scattering phase shift~(\ref{eq:phase_shift}),
 we thus arrive at a relation between the phase shift and the
 corresponding energy shift:
 \be
 \label{eq:dewitt}
 E-(m_N+\omega^*)=-{1\over\pi}\delta(\omega^*)\Delta\omega\;,
 \ee
 where $E$ is the exact energy eigenvalue perturbed from
 $(m_N+\omega^*)$.
 This result was first obtained by DeWitt long
 time ago~\cite{deWitt56:finite}. It is in fact a quite
 general result which can be derived from formal scattering theory.

 Let us now come to the discussion of the spectral weights.
 According to Eq.~(\ref{eq:spectral_weight1}), the spectral
 weights are related to the derivative of the function
 $F(x)$ evaluated at $x=E_n-m_N$. Taking the derivative
 of Eq.~(\ref{eq:Fbelow}) and Eq.~(\ref{eq:Fabove}) and
 using DeWitt's relation~(\ref{eq:dewitt}) we get:
 \makeatletter
 \let\@@@alph\@alph
 \def\@alph#1{\ifcase#1\or \or $'$\or $''$\fi}\makeatother
 \begin{subnumcases}
 {F'(x)=}
 \phi'(x), &$x < \mu$, \label{eq:Fpbelow}\\
 \phi'(x)+
 {\Gamma'(x)\over 2}\cot \delta(x)
 +{\pi\over 2}{\Gamma(x)\over \Delta\omega}
 \csc^2\delta(x), &$x\ge\mu$.
 \label{eq:Fpabove}
 \end{subnumcases}
 \makeatletter\let\@alph\@@@alph\makeatother
 It then becomes clear
 that, for eigenvalues below the threshold, the spectral weight will
 contain almost no volume dependence when the volume is
 sufficiently large:
 \be
 W_n\simeq {1\over 1+ \phi'(E_n-m_N)}\;.
 \ee
 In the Lee model, this can only happen when the
 $V$ particle is below the threshold and thus is stable.
 For eigenvalues above the threshold, however,
 the last term in Eq.~(\ref{eq:Fpabove}) is clearly
 proportional to the volume $\Omega$. As a consequence, the
 corresponding spectral weight is proportional to $1/\Omega$,
 provided the energy level is above the threshold.
 It is interesting to note that, if we take $x$ to
 be at the location of the resonance, i.e.
 $x=E_n-m_N=M^{(\delta)}-m_N$, the spectral weight is:
 \be
 \label{eq:WatR}
 W\simeq {1\over 1+\phi'(x)
 +{\pi\over 2}{\Gamma(x)\over \Delta\omega}}
 \propto {1\over 1+{\pi\over 2}{\Gamma_R(x)\over \Delta\omega}}\;,
 \ee
 where $\Gamma_R(x)\equiv \Gamma(x)/[1+\phi'(x)]$ is the
 {\em renormalized} (physical) width of the resonance~\cite{schweber:qft}.
 As was pointed out at the beginning of this
 section, all of the above discussion assumes that
 the volume is large enough. To be more precise,
 Eq.~(\ref{eq:Fpabove}) shows that, this requires
 $\Gamma_R(x)/\Delta\omega(x) \gg 1$.
 Another equivalent form for this condition is,
 combining Eq.~(\ref{eq:level-spacing}):
 \be
 \Gamma_R(x)g(x)\Omega \gg 1 \;.
 \ee
 A resonance satisfying this inequality
 is called a {\em broad} resonance. If this condition is satisfied,
 then the spectral weight~(\ref{eq:WatR}) behaves like:
 $W(x)\simeq 1/(\Gamma_R(x)g(x)\Omega)$.
 The physical meaning of the condition $\Gamma_R(x)/\Delta\omega(x) \gg 1$
 is very clear. A resonance can be considered as
 broad if its width is much larger than the
 typical level spacing in the finite box. That is
 to say, if the volume is such that within the peak of
 the resonance there are many available scattering states
 that the resonance can decay into, then the resonance is a broad one
 and the corresponding spectral weight for this resonance will
 exhibit typical two-particle state
 behavior, namely it is proportional to $1/\Omega$.
 In the opposite limit,
 \be
 \Gamma_R(x)/\Delta\omega(x) \ll 1\;,
 \;\; {\mbox{or: }}
 \Gamma_R(x)g(x)\Omega \ll 1 \;.
 \ee
 the spectral weight~(\ref{eq:WatR}) behaves like that of
 a stable single particle.
 A resonance satisfying this inequality is therefore called
 {\em infinitely narrow}. Only in this limit does a
 resonance look like a single particle as far as the
 volume dependence for the spectral weight is concerned.
 If the width and the volume are such that:
 \be
 \Gamma_R(x)g(x)\Omega \sim 1\;,
 \ee
 then the resonance is neither broad nor infinitely narrow
 and the spectral weight~(\ref{eq:WatR}) for the resonance will
 also be different from both single and two-particle spectral weights.

 \section{Discussions and conclusions}
 \label{sec:conclude}

 In this paper, we have studied the volume dependence of
 spectral weight of an unstable particle within the Lee model.
 It is shown that if the $V$-particle is stable or unstable but narrow,
 the volume dependence of the particle indeed behaves like a single
 particle, namely it is almost volume independent. However, if
 the $V$-particle is unstable and the width is large,
 then the volume dependence of its spectral weight
 exhibits two-particle properties, i.e. proportional to $1/\Omega$,
 reflecting the fact that all asymptotic states
 are two-particle scattering states.
 Thus, when the resonance changes from narrow to broad, the
 volume dependence of its spectral weight also undergoes
 a transition from a single-particle behavior to a multi-particle
 behavior. This transition can be computed exactly within the Lee
 model. The condition for a broad (or a infinitely narrow) resonance is
 also given. In real lattice QCD calculations, the criteria for a resonance
 being regarded as narrow or broad will depend on the specific problem being
 studied although the qualitative feature should remain the same.

 As an example, let us estimate the ratio $\Gamma_R/\Delta\omega$
 in lattice calculations on low-energy pion-pion scattering.
 We use this as an example because pion-pion scattering exhibits
 both a broad resonance in the scalar channel and a relatively
 narrow resonance in the vector channel.
 Recently, CP-PACS collaboration has computed the width
 of the $\rho$ resonance in the vector channel~\cite{CPPACS07:rho_decay}
 using $N_f=2$ dynamical Wilson fermion lattices of size
 $12^3\times 24$ with the lattice spacing given by: $1/a=0.92$ GeV.
 The simulation was done at $m_\pi/m_\rho=0.41$ which
 translates into pion mass of about $0.32$ GeV in physical unit.
 It is then estimated that the first and second $\Delta \omega$ to
 be about $0.5$ and $0.35$ GeV which is
 larger than the $\rho$ meson width (about $0.15$ GeV).
 Therefore, we expect that in this scenario, the $\rho$ meson
 behaves more like a narrow resonance. Indeed, the authors in
 Ref.~\cite{CPPACS07:rho_decay} have found consistent result
 for the mass of the $\rho$ meson using two different methods:
 one using the naive vector meson time correlation function,
 the other by fitting the phase shifts near the resonance.
 Note that the typical level spacing depends on
 the physical size of the volume. Since the largest physical size
 used in present lattice simulations are in the range of a few
 fermi, we expect that the typical level spacing are usually
 larger than the width of the hadron in most cases.
 An exceptional case might be the very broad $\sigma$ resonance
 in two-pion systems in the scalar channel.
 In Ref.~\cite{KFLiu06:sigma}, a
 quenched studied is performed and a single particle behavior is
 found for a scalar state in this channel. They used $16^3\times 28$ lattices
 with $a=0.2$fm and lowest pion mass is around $0.182$ GeV. The first two level
 spacings for the two pion states in this calculation are estimated to be:
 $0.45$ and $0.27$ GeV which are comparable (or somewhat smaller)
 than the expected physical width of the $\sigma$. Of course, it is difficult to
 draw definite conclusions by this naive estimate. Further studies have to
 be carried out to clarify the situation.

 To conclude, by studying the volume dependence of
 the spectral weight in a simple model, we show how
 the volume dependence of the spectral weight changes from
 single-particle to multi-particle behavior as
 the width of the resonance is getting broad.
 It is found that the ratio of its physical width $\Gamma_R$
 to the typical level spacing $\Delta\omega$ in the finite box
 controls this transition. Note that this ratio usually can be
 estimated before the simulation is actually performed, assuming
 the physical width of the resonance is known.
 We also demonstrate this by estimating this ratio for the case
 of pion-pion scattering in recent lattice calculations.
 Although studied in a simple model,
 we think that the lessons learned from the model is also
 relevant and helpful for realistic lattice simulations
 on unstable particles in QCD.

 \section*{Acknowledgments}

 The author would like to thank Prof. K.F.~Liu from
 University of Kentucky, Dr. J.P.Ma from ITP, Academia Sinica,
 Dr. Y.~Chen from IHEP, Academia Sinica,
 Prof. H.~Q.~Zheng, Prof. S.~H.~Zhu and
 Prof. S.~L.~Zhu from Peking University for valuable discussions.


\begin{thebibliography}{10}

\bibitem{luscher86:finiteb}
M.~L{\"u}scher.
\newblock Volume dependence of the energy spectrum in massive quantum field
  theories. 2. scattering states.
\newblock {\em Commun. Math. Phys.}, 105:153, 1986.

\bibitem{luscher90:finite}
M.~L{\"u}scher and U.~Wolff.
\newblock How to calculate the elastic scattering matrix in two-dimensional
  quantum field theories by numerical simulation.
\newblock {\em Nucl. Phys. B}, 339:222, 1990.

\bibitem{luscher91:finitea}
M.~L{\"u}scher.
\newblock Two particle states on a torus and their relation to the scattering
  matrix.
\newblock {\em Nucl. Phys. B}, 354:531, 1991.

\bibitem{luscher91:finiteb}
M.~L{\"u}scher.
\newblock Signatures of unstable particles in finite volume.
\newblock {\em Nucl. Phys. B}, 364:237, 1991.

\bibitem{chuan04:asymmetric}
X.~Li and C.~Liu.
\newblock Two particle states in an asymmetric box.
\newblock {\em Phys. Lett. B}, 587:100, 2004.

\bibitem{chuan04:asymmetric_long}
X.~Feng, X.~Li, and C.~Liu.
\newblock Two particle states in an asymmetric box and the elastic scattering
  phases.
\newblock {\em Phys. Rev. D}, 70:014505, 2004.

\bibitem{chuan05:2channel}
Chuan~Liu Song~He, Xu~Feng.
\newblock Two particle states and the $s$-matrix elements in multi-channel
  scattering.
\newblock {\em JHEP}, 0507:011, 2005.

\bibitem{CPPACS03:pipi_phase}
S.~Aoki et~al.
\newblock I=2 pion scattering phase shift with wilson fermions.
\newblock {\em Phys. Rev. D}, 67:014502, 2003.

\bibitem{CPPACS04:pipi_phase_unquench}
T.~Yamazaki et~al.
\newblock I=2 $\pi\pi$ scattering phase shift with two flavors of o(a) improved
  dynamical quarks.
\newblock {\em Phys. Rev. D}, 70:074513, 2004.

\bibitem{savage06:pipi}
Silas~R. Beane, Paulo~F. Bedaque, Kostas Orginos, and Martin~J. Savage.
\newblock I=2 pi-pi scattering from fully-dynamical mixed-action lattice qcd.
\newblock {\em Phys. Rev. D}, 73:054503, 2006.

\bibitem{savage06:NN}
Silas~R. Beane, Paulo~F. Bedaque, Thomas~C. Luu, Kostas Orginos, Elisabetta
  Pallante, Assumpta Parreno, and Martin~J. Savage.
\newblock Nucleon-nucleon scattering from fully-dynamical lattice qcd.
\newblock {\em Phys. Rev. Lett.}, 97:012001, 2006.

\bibitem{savage06:piK}
S.R. Beane, P.F. Bedaque, K.~Orginos, and M.J. Savage.
\newblock Pi-k scattering in full qcd with domain-wall valence quarks.
\newblock {\em Phys. Rev. D}, 74:114503, 2006.

\bibitem{michael05:decay}
C.~Michael.
\newblock Hadronic decays.
\newblock {\em PoS LAT2005}, page 008, 2005.

\bibitem{michael06:hybrid_decay}
C.~McNeile and C.~Michael.
\newblock Decay width of light quark hybrid meson from the lattice.
\newblock {\em Phys. Rev. D}, 73:074506, 2006.

\bibitem{fiebig06:hybrid_decay}
M.~S. Cook and H.~R. Fiebig.
\newblock Exotic meson decay widths using lattice qcd.
\newblock {\em Phys. Rev. D}, 74:034509, 2006.

\bibitem{CPPACS07:rho_decay}
K-I. Ishikawa N. Ishizuka K. Kanaya Y. Kuramashi Y. Namekawa M. Okawa K. Sasaki
  A. Ukawa T.~Yoshi¨¦ S.~Aoki, M.~Fukugita.
\newblock Lattice qcd calculation of the $\rho$ meson decay width.
\newblock {\em Phys. Rev. D}, 76:094506, 2007.

\bibitem{KFLiu04:pentaquark}
N.~Mathur, F.X. Lee, A.~Alexandru, C.~Bennhold, Y.~Chen, S.J. Dong, T.~Draper,
  I.~Horvath, K.F. Liu, S.~Tamhankar, and J.B. Zhang.
\newblock A study of pentaquarks on the lattice with overlap fermions.
\newblock {\em Phys. Rev. D}, 70:074508, 2004.

\bibitem{Takahashi:2005uk}
Toru~T. Takahashi, Takashi Umeda, Tetsuya Onogi, and Teiji Kunihiro.
\newblock {Search for the possible S = +1 pentaquark states in quenched lattice
  QCD}.
\newblock {\em Phys. Rev.}, D71:114509, 2005.

\bibitem{Csikor:2005xb}
F.~Csikor, Z.~Fodor, S.~D. Katz, T.~G. Kovacs, and B.~C. Toth.
\newblock {A comprehensive search for the Theta+ pentaquark on the lattice}.
\newblock {\em Phys. Rev.}, D73:034506, 2006.

\bibitem{Ishii:2004qe}
N.~Ishii et~al.
\newblock {Penta-quark baryon in anisotropic lattice QCD}.
\newblock {\em Phys. Rev.}, D71:034001, 2005.

\bibitem{Lee54:model}
T.D. Lee.
\newblock Some special examples in renormalizable field theory.
\newblock {\em Phys. Rev.}, 95:1329, 1954.

\bibitem{Glaser55:unstable}
V.~Glaser and G.~K{\"a}ll{\'e}n.
\newblock A model of unstable particle.
\newblock {\em Nucl. Phys.}, 2:706, 1957.

\bibitem{schweber:qft}
S.S. Schweber.
\newblock {\em An Introduction to Relativistic Quantum Field Theory}.
\newblock Evanston, Ill.: Row, Peterson, 1961.

\bibitem{deWitt56:finite}
B.S. DeWitt.
\newblock Transition from discrete to continuous spectra.
\newblock {\em Phys. Rev.}, 103:1565, 1956.

\bibitem{KFLiu06:sigma}
Nilmani Mathur, A.~Alexandru, Y.~Chen, S.J. Dong, T.~Draper, I.Horvath, F.X.
  Lee, K.F. Liu, S.~Tamhankar, and J.B. Zhang.
\newblock Scalar mesons $a_0(1450)$ and $\sigma(600)$ from lattice qcd.
\newblock {\em Phys. Rev. D}, 76:114505, 2007.

\end{thebibliography}

 \end{document}